\documentclass[aps,a4paper,10pt,twocolumn,nofootinbib,floatfix]{revtex4} 
\usepackage{datetime}
\usepackage{amsmath}
\usepackage{amsfonts}
\usepackage{amssymb}
\usepackage{mathrsfs}
\usepackage[mathscr]{euscript}
\usepackage[dvips]{graphicx}
\usepackage{fancyhdr}
\usepackage[hypertex=true]{hyperref} 
\usepackage{makeidx}
\usepackage{colordvi}

\def\overstrike#1#2{{\setbox0\hbox{$#2$}\hbox to \wd0{\hss
    $#1$\hss}\kern-\wd0\box0}}

\newcommand{\convol}{\star}

        \DeclareMathOperator{\grad}{\nabla}

\newcommand{\XDOI}[1]{\href{http://dx.doi.org/#1}{doi:#1}}

\newdateformat{yymmdddate}{\THEYEAR/\twodigit{\THEMONTH}/\twodigit{\THEDAY}}

\begin{document}

\title{Directional pulse propagation in beam, rod, pipe, and disk geometries}

\author{P. Kinsler}
\email{Dr.Paul.Kinsler@physics.org}

\affiliation{
  Blackett Laboratory, Imperial College London,
  Prince Consort Road,
  London SW7 2AZ,
  United Kingdom.}

\begin{abstract}

I derive directional wave equations useful for 
 pulses propagating in 
 beam, 
 rod,
 pipe, 
 and disk geometries by using a cylindrical coordinate system; 
 the scheme works equally well for either long multi-cycle or single-cycle
 ultrashort pulses.
This is achieved by means of a factorization procedure
 that conveniently generates
 exact bi-directional and first order wave equations
 after the selection of propagation direction -- 
 either axial, 
 radial, or even angular.
I then discuss how to reduce these to a uni-directional form, 
 and discuss the necessary approximation, 
 which is essentially a paraxial approximation
 as appropriately generalized to the specific geometry.

\end{abstract}

\lhead{\includegraphics[height=5mm,angle=0]{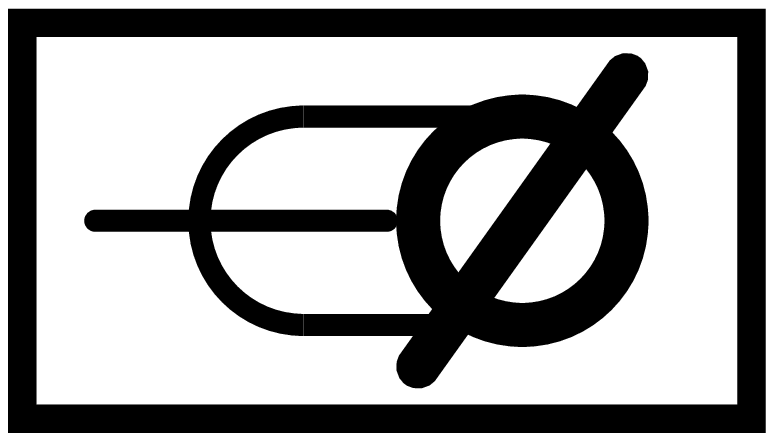}~~FBRAD}
\chead{Directional propagation in cylindrical geometries}
\rhead{
\href{mailto:Dr.Paul.Kinsler@physics.org}{Dr.Paul.Kinsler@physics.org}\\
\href{http://www.kinsler.org/physics/}{http://www.kinsler.org/physics/}
}

\date{\today}
\maketitle
\thispagestyle{fancy}

\section{Introduction}
\label{S-Introduction}

Directional decompositions of wave equations
 \cite{Kinsler-2010pra-fchhg,Genty-KKD-2007oe,Kinsler-RN-2005pra,Kolesik-MM-2002prl,Kinsler-2012arXiv-fbacou}
 are a powerful method of developing pulse propagation equations
 valid down to the ultrashort and few cycle regime.
Done correctly, 
 these provide exact bi-directional forms, 
 which can be systematically approximated in a ``slow evolution'' limit
 into a uni-direction form of great practical use.
Usually, 
 the emphasis is on the role of nonlinearity or dispersion, 
 and on modelling the propagation of beams or pulses
 in a linear geometry.

A feature of the decomposition is the choice of a reference evolution, 
 containing as much of the detail of the exact propagation as possible, 
 with the rest left as a (hopefully) perturbative ``residual'' term
 that couples the forward and backward waves.
Fortunately, 
 even if the residual is not that weak, 
 the backward contribution is exceedingly poorly phase matched, 
 which greatly extends the validity of the approximation.
Nonlinearity is typically left as a residual term, 
 as is diffraction.

Here my intention is to take an alternative path, 
 and provide the basics of
 bi- and uni-directional propagation models
 for non-Cartesian geometries,
 although at first I restrict myself to the cylindrical case
 relevant to beams, rods, and disks; 
 others can be easily generated as required.
In particular, 
 although the result of factorization for directed axial propagation
 will likely look familiar, 
 the radial and angular cases are more interesting.
In particular, 
 such propagation models have potential applications 
 for disk, ring, and other whispering galley optical resonators
 \cite{Mirsky-1965jasa,Gorodetsky-SI-1996ol,Armani-KSV-2003n,Absil-HLWJH-2000ptl,Gayral-GLDMP-1999apl}.
My focus on geometry rather than dispersion or nonlinearity, 
 means that it is the diffraction terms that are of more interest
 than nonlinearity and dispersion; 
 for these the ``slow evolution'' criteria amounts
 to a generalised paraxial approximation.
Although the overt focus is on optical pulse propagation, 
 since the Helmholtz wave equation used as a starting point 
 is useful in many fields (e.g. for acoustic pressure waves), 
 the results here have potential for wider application.



\section{Theory}
\label{S-basic}

Most optical pulse problems consider 
 a uniform and source free dielectric medium.
In such cases a good starting point 
 is the second order wave equation, 
 which results from the substitution of the $\grad \times \vec{H}$
 Maxwell's equation into the $\grad \times \vec{E}$ one
 in the source-free case
 (see e.g. \cite{Agrawal-NFO}).
Further, 
 assuming linearly polarized pulses, 
 we can use a scalar form.
Defining 
 $\grad^2 = \partial_x^2 + \partial_y^2 + \partial_z^2$
 and $\partial_a \equiv \partial / \partial a$, 
 we can write the wave equation as
~
\begin{align}
  \left[
    \grad^2
   -
    \frac{1}{c^2}
    \partial_t^2
    \epsilon \convol \mu \convol 
  \right]
  E(t)
&=
  \mathscr{Q}
.
\label{eqn-basic-nabla2E}
\end{align}
Here I have suppressed the spatial coordinates for notational simplicity; 
 in fact we have $E(t) \equiv E(t,\vec{r})$ 
 and the total polarization $\mathscr{Q}(\Vec{E},t,\vec{r})$; 
 also $\vec{r} = (x,y,z)$.
Note that a full expansion of the various possible components of $\mathscr{Q}$
 is given in \cite{Kinsler-2010pra-fchhg},
 but in summary it can contain nonlinearity,
 dispersion, 
 and free current effects -- 
 and potentially even magnetic nonlinearity. 
It can even allow for non-Helmholtz behaviour, 
 such as that present in some acoustic wave models
 \cite{Kinsler-2012arXiv-fbacou}; 
 and if adapted can generate temporally propagated wave equations
 instead of the spatially propagated ones
 described here \cite{Kinsler-2012arXiv-fbacou}.

Of the potential complications, 
 we here include the isotropic linear material response terms
 in a reference wavevector 
 $k^2(\omega) = \epsilon(\omega)\mu(\omega) \omega^2$.
Thus, 
 in the frequency domain, 
 we can write
~
\begin{align}
  \left[
    \grad^2
   -
    k^2(\omega)
  \right]
  E(t)
&=
 -
  \mathscr{Q}
.
\label{eqn-basic-nabla2Ew-raw}
\end{align}

In most descriptions of pulse propagation we will want to chose a 
 specific propagation direction 
 and then denote the orthogonal components 
 as transverse behaviour.
Often this process uses Cartesian $x, y, z$ coordinates
 (see e.g. \cite{Kinsler-2010pra-fchhg},
 but here I show how directional techniques 
 can be applied in alternative geometries.


I now factorize the wave equation, 
 a process which, 
 while used in optics for some time \cite{Blow-W-1989jqe}
 has only recently been used to its full potential
 \cite{Ferrando-ZCBM-2005pre,Genty-KKD-2007oe,Kinsler-2007josab,Kinsler-2012arXiv-fbacou}.
Given a wave equation of the form 
~
\begin{align}
  \left[
    \partial_z^2
   +
    K^2
  \right]
  E
&=
 -
  \bar{\mathscr{Q}}
,
\label{eqn-basic-nabla2Ew}
\end{align}
 with $\bar{\mathscr{Q}}$ now also including
 the non-$\partial_z^2$ derivative terms,
 we can see that the LHS of eqn. \eqref{eqn-basic-nabla2Ew}
 is a simple sum of squares which might be factorized, 
 indeed this is what was done in a somewhat ad hoc fashion 
 by Blow and Wood in 1989 \cite{Blow-W-1989jqe}.
Since the factors are just $\partial_z \mp \imath K$, 
 we can see that each (by itself) would generate
 a forward directed wave equation, 
 and the other a backward one.
Leaving basic mathematical detail to the appendix, 
 a rigorous factorization procedure
 \cite{Ferrando-ZCBM-2005pre,Kinsler-2010pra-fchhg}
 allows us to define a pair of counter-propagating Greens functions, 
 and so divide the second order wave equation
 into a pair of coupled \emph{counter-propagating} first order ones.

Counter-propagating wave equations suggest counter propagating fields, 
 so I split the electric field up accordingly into forward ($E^+$) 
 and backward ($E^-$) parts, 
 with $E = E^+ + E^-$.
The coupled first order wave equations are
~
\begin{align}
    \partial_z
  E^\pm
&=
 \pm
    \imath
    K
  E^\pm
 \quad
 \pm
  \frac{\imath \bar{\mathscr{Q}}}{2K}
.
\label{eqn-bi-dzE}
\end{align}
The RHS now falls into two parts, 
 which I term the underlying and residual parts \cite{Kinsler-2009pra}.
First, 
 there is the $\imath K E^\pm$ term
 that, 
 by itself, 
 will describe plane-wave like propagation in the simplest cases.
Second, 
 the remaining part $\propto \bar{\mathscr{Q}}$
 which can be called ``residual'' terms.
These residual contributions, 
 here containing the transverse derivatives $\partial_x^2+\partial_y^2$,
 account for the discrepancy between the true propagation
 and the underlying propagation.
Although here we might hope that 
 this residual component is only a weak perturbation, 
 the theory presented here is valid for \emph{any} strength.
This wide validity is of course very advantageous, 
 however note that this approach is most useful in the uni-directional limit, 
 i.e. when the residual terms \emph{are} small
 in addition
 to being poorly phase matched \cite{Kinsler-2010pra-fchhg}\footnote{
  If the forward field has a wave vector $k_0$ 
   evolving as $\exp(+\imath k_0 z)$, 
   the generated backward component will evolve as $\exp(-\imath k_0 z)$.
  This gives a very rapid relative oscillation $\exp(-2\imath k_0 z)$, 
   which will quickly average to zero.}

Here I only consider the effects of diffraction in any detail; 
 other effects are not the specific subject of this work, 
 and due to their lesser significance (here)
 are assumed to be incorporatd in the residual term $\mathscr{Q}$.

\section{Beams, rods, pipes, and disks}
\label{S-cylindrical}

The cylindrical geometry is perhaps the most likely
 to give useful results, 
 as it covers not only the common case of a light beam of circular profile, 
 but also propagation around the edge of a disk resonator.
Here the coordinates are the axial $z$, 
 the radial $\rho$
 and an angle $\theta$.
In the rest of this section, 
 I choose each in turn as the direction for the underlying propagation, 
 although by far the most common case is the axial 
 ``axi-symmetric'' 
 case relevant for the typical light beam.

To proceed we will need an expression for the Laplacian $\grad^2$
 in cylindrical coordinates, 
 which is just the usual expression
~
\begin{align}
  \grad^2 E
&=
  \frac{1}{\rho}
  \partial_\rho
  \left(
    \rho
    \partial_\rho
    E
  \right)
 +
  \frac{1}{\rho^2}
  \partial_\phi^2
    E
 +
  \partial_z^2
    E
,
\label{eqn-Laplacian-cyl}
\end{align}
 where now $E \equiv E(z,\rho,\phi;\omega)$.
From this point we only need choose a primary propagation axis
 according to our interests, 
 and proceed from there.
In the following, 
 I consider each possible choice in turn.

\subsection{Axial}
\label{S-axial-rho}

This axial case is suitable for the common case
 of free-space beam propagation, 
 or that along a slowly changing rod
 or circular waveguide, 
 such as a tapered optical fibre \cite{Agrawal-NFO}.
This is because we would expect angular variation 
 and radial variation to be small and/or only slowly varying.
Note that axial propagation 
 along the $z$ coordinate has the nice feature
 that the propagation coordinate 
 is translationally invariant along itself; 
 i.e. we do not have to care where ``$z=0$'' is.

\begin{figure}
\includegraphics[width=0.80\columnwidth,angle=0]{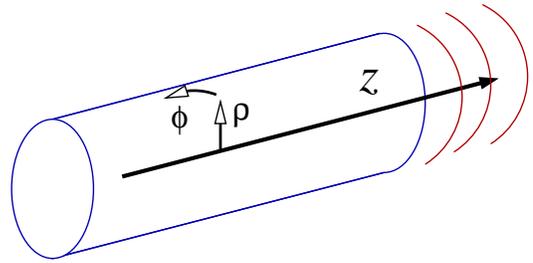}
\caption{Longitudinal propagation in cylindrical coordinates: 
 here the pulse propagates towards larger $z$, 
 whilst transverse variation occurs in the radial $\rho$
 and angular $\phi$ directions.
}
\label{fig-cyl-z}
\end{figure}

As indicated on fig. \ref{fig-cyl-z}, 
 we choose the propagation direction along
 the $z$ axis of the cylindrical coordinates, 
 with radial coordinate $\rho$
 and angular coordinate $\theta$
 to account for any transverse variation.
This contrast with models 
 (e.g. \cite{Kinsler-2010pra-fchhg})
 which use the Cartesian $x, y$ as transverse coordinates, 
 although of course the use of radial transverse coordinates
 is far from unknown [refs].
We rewrite the wave equation \eqref{eqn-basic-nabla2Ew} 
 to focus on $z$-propagation, 
 and demote radial and angular effects to the status of residual terms, 
 resulting in
~
\begin{align}
  \left[
    \partial_z^2 
   +
    K^2
  \right]
  E
&=
 -
  \mathscr{Q}
 -
  \frac{1}{\rho}
  \partial_\rho
   \rho
  \partial_\rho
  E
 -
  \frac{1}{\rho^2}
  \partial_\phi^2
  E
,
\label{eqn-cyl-d2zE}
\end{align}
where the total wavevector is given by $K^2 = n^2 (\omega) \omega^2 / c^2$.
Factorizing gives us
~
\begin{align}
    \partial_z
  E^\pm
&=
 \pm
    \imath
    K
  E^\pm
 \pm
  \frac{\imath \mathscr{Q}}{2K}
 \pm
  \frac{\imath }{2\rho K}
  \partial_\rho
   \rho
  \partial_\rho
  \left[ E^+ + E^- \right]
\nonumber\\
& \qquad 
 \pm
  \frac{\imath }{2\rho^2 K}
  \partial_\phi^2
  \left[ E^+ + E^- \right]
.
\label{eqn-bicyl-dzE}
\end{align}
Here, 
 in addition to the $\mathscr{Q}$ residual term, 
 we have two additional coordinate-based residual terms.
The first is that which gives radial diffraction, 
 and the second angular diffraction.
Both of these appear to have potential singularities at $\rho=0$, 
 but this is a coordinate effect -- 
 the singularity is not present in Cartesian coordinates.
Thus, 
 $E$ will typically be smooth enough so that this
 will not cause pathological difficulties.

Assuming both the radial and angular diffraction terms are small, 
 we can decouple the $E^+$ and $E^-$ fields 
 as described and justified in more detail in \cite{Kinsler-2010pra-fchhg}.
For this to hold, 
 we need all the residual terms on the RHS to be much smaller
 than the leading $KE^\pm$ term, 
 i.e.
~
\begin{align}
  \left| \mathscr{Q} \right|    
 \ll   
  \left| 2K^2 E^\pm \right|, 
\\
  \left| \partial_\rho \rho \partial_\rho \left(E^+ + E^- \right) \right|    
 \ll   
  \left| 2K^2 \rho E^\pm \right|, 
\\
  \left| \partial_\phi^2 \left(E^+ + E^- \right) \right|    
\ll   
  \left| 2K^2  \rho^2 E^\pm \right|
.
\end{align}
These being sufficently well satisfied, 
 we can approximate eqn. \eqref{eqn-bicyl-dzE}
 to get the uni-directional wave equation 
 for propagation in a beam or rod which is
~
\begin{align}
    \partial_z
  E^\pm
&=
 \pm
    \imath
    K
  E^\pm
 \pm
  \frac{\imath \mathscr{Q}}{2K}
 \pm
  \frac{\imath }{2\rho K}
  \partial_\rho
   \rho
  \partial_\rho
  E^\pm
 \pm
  \frac{\imath }{2\rho^2 K}
  \partial_\phi^2
  E^\pm
.
\label{eqn-unicyl-dzE}
\end{align}

\subsection{Radial}
\label{S-cylindrical-rho}

The radial case might be applied to the case where a wire or point source
 is radiating outwards into a cylinder or disk; 
 or perhaps the reverse situation with converging fields.
Alternatively, 
 it might be useful when approaching the far-field, 
 where part of an expanding wavefront
 enters some area of interest.
Unlike the axial case where the absolute location $z=0$
 was unimportant, 
 here the coordinate centre at $\rho=0$ is fixed.

\begin{figure}
\includegraphics[width=0.80\columnwidth,angle=0]{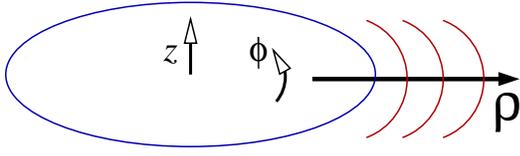}
\caption{Radial propagation in cylindrical coordinates: 
 here the pulse propagates towards larger $\rho$, 
 whilst transverse variation occurs in the axial $z$
 and angular $\phi$ directions.
}
\label{fig-cyl-rho}
\end{figure}

As shown in fig. \ref{fig-cyl-rho}, 
 we choose the propagation direction along
 the $\rho$ radial coordinate, 
 with the axial $z$ and angular $\theta$ coordinates
 to account for any transverse variation.
We rewrite the wave equation \eqref{eqn-basic-nabla2Ew} 
 to focus on $\rho$-propagation, 
 and demote axial and angular effects to the status of residual terms, 
 resulting in
~
\begin{align}
  \left[
  \frac{1}{\rho}
  \partial_\rho
   \rho
  \partial_\rho
   +
    K^2
  \right]
  E
&=
  \mathscr{Q}
 -
    \partial_z^2 
  E
 -
  \frac{1}{\rho^2}
  \partial_\phi^2
  E
,
\label{eqn-cyl-d2rE}
\end{align}
where $E \equiv E(z,\rho,\phi;\omega)$, 
 and the total wavevector is given by $K^2 = n^2(\omega) \omega^2 / c^2$.
Now since 
\begin{align}
  \rho^{-1}
  \partial_\rho
    \rho
    \partial_\rho
      E
&=
  \rho^{-1}
  \partial_\rho
  \left[
    \partial_\rho \rho E
   -
    E
  \right]
,
\label{eqn-cyl-rident}
\end{align}
 then with $F_\rho=\rho E$, 
 we get 
~
\begin{align}
  \left[
   \partial_\rho^2
   +
    K^2
  \right]
  F_\rho
&=
 -
  \rho
  \mathscr{Q}
 -
  \rho
    \partial_z^2 
  F_\rho
 -
  \frac{1}{\rho^2}
  \partial_\phi^2
  F_\rho
 +
    \partial_\rho
  \frac{F_\rho}{\rho}
.
\label{eqn-cyl-d2rE-alt}
\end{align}
Factorizing gives us
~
\begin{align}
    \partial_\rho
  F_\rho^\pm
&=
 \pm
    \imath
    K
  F_\rho^\pm
 \pm
  \frac{\imath \rho \mathscr{Q}}{2K}
 \pm
  \frac{\imath \rho}{2K}
  \partial_z^2 
  \left[ F_\rho^+ + F_\rho^- \right]
\nonumber\\
& \qquad 
 \pm
  \frac{\imath}{2 \rho^2 K}
  \partial_\phi^2 
  \left[ F_\rho^+ + F_\rho^- \right]
 \mp
  \frac{\imath}{2 K}
  \partial_\rho
  \frac{ F_\rho^+ + F_\rho^- }{\rho}
.
\label{eqn-bicyl-drE}
\end{align}

One feature of this is that the RHS has a residual term
 containing a $\partial_\rho$ derivative, 
 which was generated when moving from the field $E$
 to the radially scaled version $F_\rho$.
We could therefore move this to the LHS now, 
 but for simplicity I delay this adjustment until after the 
 uni-directional approximation is made.

Here, 
 in addition to the generic $\mathscr{Q}$ residual term, 
 we have three additional coordinate-based residual terms.
The first is that which gives axial diffraction, 
 and the second angular diffraction.
The third arrives as a result of eqn. \eqref{eqn-cyl-rident}, 
 and acts as a radial drift.
The second and third 
 of these appear to have potential singularities at $\rho=0$, 
 but this is a coordinate effect -- 
 the singularity is not present in Cartesian coordinates.
Thus, 
 $F$ will typically be smooth enough so that this
 will not cause pathological difficulties.

Assuming both the axial and angular diffraction terms are small, 
 we can decouple the $E^+$ and $E^-$ fields 
 as described and justified in more detail in \cite{Kinsler-2010pra-fchhg}.
For this to hold, 
 we need all the residual terms on the RHS to be much smaller
 than the leading $KE^\pm$ term, 
 i.e.
~
\begin{align}
  \left| \rho \mathscr{Q} \right|    
 \ll   
  \left| 2K^2 F_\rho^\pm \right|, 
\\
  \left| \rho \partial_z^2 \left(F_\rho^+ + F_\rho^- \right) \right|    
 \ll   
  \left| 2K^2 F_\rho^\pm  \right|, 
\\
  \left| \partial_\phi^2 \left(F_\rho^+ + F_\rho^- \right) \right|    
\ll   
  \left| 2K^2  \rho^2 F_\rho^\pm \right|
\\
  \left| \partial_\rho \rho^{-1} \left(F_\rho^+ + F_\rho^- \right) \right|    
\ll   
  \left| 2K^2 F_\rho^\pm \right|
.
\end{align}
Of these, 
 note in particular the last one, 
 where we can see that we will need to be away from the origin
 for it to hold -- 
 as indeed might be expected on physical grounds.
These being sufficently well satisfied, 
 we can approximate eqn. \eqref{eqn-bicyl-drE}
 to get the uni-directional wave equation 
 for outward or inward radial propagation, 
 which is
~
\begin{align}
    \partial_\rho
  F_\rho^\pm
&=
 \pm
    \imath
    K
  F_\rho^\pm
 \pm
  \frac{\imath \rho \mathscr{Q}}{2K}
 \pm
  \frac{\imath \rho}{2K}
  \partial_z^2 
  F_\rho^\pm
\nonumber\\
& \qquad 
 \pm
  \frac{\imath}{2 \rho^2 K}
  \partial_\phi^2 
  F_\rho^\pm
 \mp
  \frac{\imath}{2 K}
  \partial_\rho
  \frac{F_\rho^\pm}{\rho}
.
\label{eqn-unicyl-drE}
\end{align}

Now we can combine the two $\partial_\rho$ derivatives, 
 to get
~
\begin{align}
    \partial_\rho
  \left(
   1 \pm \frac{1}{2K\rho}
  \right)
  F_\rho^\pm
&=
 \pm
    \imath
    K
  F_\rho^\pm
 \pm
  \frac{\imath \rho \mathscr{Q}}{2K}
 \pm
  \frac{\imath \rho}{2K}
  \partial_z^2 
  F_\rho^\pm
 \pm
  \frac{\imath}{2 \rho^2 K}
  \partial_\phi^2 
  F_\rho^\pm
.
\label{eqn-unicyl-drE--v2}
\end{align}

\subsection{Angular}
\label{S-cylindrical-phi}

Angular propagation is relevant where the light is propagating around 
 some kind of circular waveguide, 
 such as in a whispering-gallery (disk) waveguide, 
 although it could also be applied to a helical waveguide.
The restriction to waveguides results from the fact 
 that without some confining structure, 
 light will travel in a straight line, 
 and so would only only be nearly angular for a brief interval
 at closest approach to the coordinate origin.
The angular case has the nice feature that the propagation coordinate (?)
 is translationally invariant along itself (around the origin); 
 i.e. we do not have to care where ``$\theta=0$'' is.

\begin{figure}
\includegraphics[width=0.80\columnwidth,angle=0]{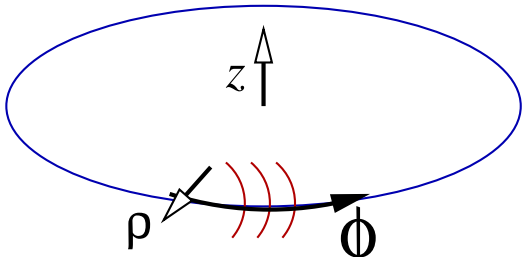}
\caption{Angular propagation in cylindrical coordinates: 
 here the pulse propagates around in positive $\phi$,
 whilst transverse variation occurs in the axial $z$
 and radial $\rho$ directions.
}
\label{fig-cyl-phi}
\end{figure}

As shown in fig. \ref{fig-cyl-phi},
 here we choose the propagation direction around
 the $\phi$ angular coordinates
 with the axial $z$ and radial $\theta$ coordinates
 to account for any transverse variation.
We rewrite the wave equation \eqref{eqn-basic-nabla2Ew} 
 to focus on $\phi$-propagation, 
 and demote axial and radial effects to the status of residual terms, 
 resulting in
~
\begin{align}
  \left[
    \frac{1}{\rho^2}
    \partial_\phi^2
   +
    K^2
  \right]
  E
&=
  \mathscr{Q}
 -
    \partial_z^2 
  E
 -
  \frac{1}{\rho}
  \partial_\rho
   \rho
  \partial_\rho
  E
,
\label{eqn-cyl-d2phiE}
\end{align}
 with  $E \equiv E(z,\rho,\phi;\omega)$, 
 and where the total wavevector is given by $K^2 = n^2(\omega) \omega^2 / c^2$.
Then with $F_\phi=\rho^2 E$, 
 we get 
~
\begin{align}
  \left[
   \partial_\phi^2
   +
    K^2
  \right]
  F_\phi
&=
  \rho^2
  \mathscr{Q}
 -
  \rho^2
    \partial_z^2 
  F_\rho
 -
  \frac{1}{\rho}
  \partial_\rho
   \rho
  \partial_\rho
  F_\rho
.
\label{eqn-cyl-d2phiE-alt}
\end{align}
Factorizing gives us
~
\begin{align}
    \partial_\phi
  F_\phi^\pm
&=
 \pm
    \imath
    \rho
    K
  F_\phi^\pm
 \pm
  \frac{\imath \rho^2 \mathscr{Q}}{2K}
 \pm
  \frac{\imath \rho^2}{2K}
  \partial_z^2 
  \left[ F_\phi^+ + F_\phi^- \right]
\nonumber\\
& \qquad 
 \pm
  \frac{\imath \rho }{2 K}
  \partial_\rho
   \rho
    \partial_\rho
    \left[ F_\phi^+ + F_\phi^- \right]
.
\label{eqn-bicyl-dphiE}
\end{align}
Here, 
 in addition to the $\mathscr{Q}$ residual term, 
 we have two additional coordinate-based residual terms.
The first is that which gives axial diffraction, 
 and the second radial diffraction; 
 although the second may be split into an alternate diffraction
 along with a radial drift by using eqn. \eqref{eqn-cyl-rident}.

Assuming both the axial and radial diffraction terms are small, 
 we can decouple the $E^+$ and $E^-$ fields 
 as described and justified in more detail in \cite{Kinsler-2010pra-fchhg}.
For this to hold, 
 we need all the residual terms on the RHS to be much smaller
 than the leading $KE^\pm$ term, 
 i.e.
~
\begin{align}
  \left| \rho \mathscr{Q} \right|    
 \ll   
  \left| 2K^2 F_\phi^\pm \right|, 
\\
  \left| \rho \partial_z^2 \left(F_\phi^+ + F_\phi^- \right) \right|    
 \ll   
  \left| 2K^2 F_\phi^\pm  \right|, 
\\
  \left| \partial_\rho \rho \partial_\rho \left(F_\phi^+ + F_\phi^- \right) \right|    
\ll   
  \left| 2K^2  F_\phi^\pm \right|
.
\end{align}
These being sufficently well satisfied, 
 we can approximate eqn. \eqref{eqn-bicyl-dphiE}
 to get the uni-directional wave equation 
 for angular propagation,
 which is
~
\begin{align}
    \partial_\phi
  F_\phi^\pm
&=
 \pm
    \imath
    \rho
    K
  F_\phi^\pm
 \pm
  \frac{\imath \rho^2 \mathscr{Q}}{2K}
 \pm
  \frac{\imath \rho^2}{2K}
  \partial_z^2 
  F_\phi^\pm
\nonumber\\
& \qquad 
 \pm
  \frac{\imath \rho }{2 K}
  \partial_\rho
   \rho
    \partial_\rho
    F_\phi^\pm
.
\label{eqn-unicyl-dphiE}
\end{align}

Since to maintain uni-directionality during this kind of angular propagation, 
 our wave must be somehow confined in a ring shaped waveguide, 
 most likely the radial terms included above will not be relevant -- 
 any radial diffraction will have been already balanced
 by the radially confining waveguide structure, 
 and the radial wave profile will match some guided mode.
In this case,  
 we can use 
~
\begin{align}
    \partial_\phi
  F_\phi^\pm
&=
 \pm
    \imath
    \rho
    K
  F_\phi^\pm
 \pm
  \frac{\imath \rho^2 \mathscr{Q}}{2K}
 \pm
  \frac{\imath \rho^2}{2K}
  \partial_z^2 
  F_\phi^\pm
,
\label{eqn-unicyl-dphiE-rholess}
\end{align}
 to propagate light pulses around a thick disk, 
 ring, 
 or pipe,
 whilst still allowing for axial diffraction.

%
\section{Conclusion}
\label{S-Conclude}

Here I have derived bi-directional factorizations
 of the Helmhotz wave equation
 in the cylindrical geometry,
 focussing on each possible choice of propagation direction in turn.
These then allow approximate uni-directional forms, 
 based on a generalized notion of paraxiality; 
 and it is these which are likely to be most useful.
These results are done in the same style as, 
 and are intended to complement existing 
 calculations done using cartesian coordinates \cite{Kinsler-2010pra-fchhg}.

One could certainly also image following this same procedure
 using other orthogonal coordinate systems\footnote{See
  e.g. \url{http://en.wikipedia.org/wiki/Coordinate_system\#List\_of\_orthogonal\_coordinate\_systems}}, 
 notably spherical-polars or parabolic coordinates.
You could also use the approach to model diffraction of ray-like
 light beams in a conformal cloak \cite{Leonhardt-2006sci} or similar; 
 providing sufficient physical motivation exists
 and the a uni-directional approximation 
 to the resulting wave propagation equations is achievable.

%



%


%
\section*{Appendix: Factorizing}
\label{S-factorize}

Here is a quick derivation of the factorization process; 
 the $z$-derivative has been converted to $\imath k$, 
 $\beta^2 = n^2 \omega^2 /c^2$,
 and the unspecified residual term is denoted $Q$.
~
\begin{align}
  \left[
   -k^2 + \beta^2
  \right]
  E
&=
 -Q
\\
  E
&=
  \frac{1}{k^2 - \beta^2}
  Q
\qquad
=
  \frac{1}{\left(k-\beta\right)\left(k+\beta\right)}
\\
&=
  \frac{-1}{2\beta}
  \left[
    \frac{1}{k+\beta}
   -
    \frac{1}{k-\beta}
  \right]
  Q
.
\end{align}
Now $(k-\beta)^{-1}$ is a forward-like propagator for the field, 
 and $(k+\beta)^{-1}$ a backward-like propagator.
Hence write $E=E^++E^-$, 
 and split the two sides up
~
\begin{align}
  E^+
 +
  E^-
&=
  \frac{-1}{2\beta}
  \left[
    \frac{1}{k+\beta}
   -
    \frac{1}{k-\beta}
  \right]
  Q
\\
  E^\pm
&=
  \frac{\pm1}{2\beta}
    \frac{1}{k\mp\beta}
  Q
\\
  \left[
   k \mp \beta
  \right]
  E^\pm
&=
 \pm
  \frac{1}{2\beta}
    \frac{1}{k\mp\beta}
  Q
\\
  \imath
  k E^\pm
&= 
 \pm
  \imath
  \beta E^\pm
 \pm
  \frac{\imath}{2\beta}
  Q
,
\end{align}
and reverting to $z$ derivatives gives us the final form
~
\begin{align}
  \partial_z
  E^\pm
&= 
 \pm
  \imath
  \beta E^\pm
 \pm
  \frac{\imath}{2\beta}
  Q
.
\end{align}

\end{document}